\documentclass[twocolumn,pr,showpacs,notitlepage,nobibnotes,nofootinbib]{revtex4-1}
\usepackage{bm}
\usepackage{graphicx}
\usepackage{amsmath}
\usepackage{amssymb}
\usepackage{eufrak}
\usepackage{color}
\newcommand{\nix}[1]{}

\usepackage{array}

\newcount\timehh  \newcount\timemm
\timehh=\time \divide\timehh by 60
\timemm=\time
\begin{document}

\title{Giant Zeeman splitting of light holes in GaAs/AlGaAs quantum wells} 
\author{M.V. Durnev, M.M. Glazov, E.L. Ivchenko}
\affiliation{Ioffe Physical-Technical Institute, RAS, 194021
  St.-Petersburg, Russia}

%\date{\today, file = \jobname.tex, printing time =
 % \number\timehh\,:\,\ifnum\timemm<10 0\fi \number\timemm}

\begin{abstract}
We have developed a theory of the longitudinal $g$ factor of light holes in 
semiconductor quantum wells. It is shown that the absolute value of
the light-hole $g$-factor can strongly exceed its value in the bulk
and, moreover, the dependence of the Zeeman splitting on magnetic field becomes
non-linear in relatively low fields. These effects are determined by
the proximity of the ground light-hole subband, $lh1$, to the first
excited heavy-hole subband, $hh2$, in GaAs/AlGaAs-type structures. The
particular calculations are performed in the framework of Luttinger Hamiltonian
taking into account both the magnetic field-induced mixing of $lh1$ and
$hh2$ states and the mixing of these states at heterointerfaces, the latter caused
by chemical bonds anisotropy. A theory of magneto-induced reflection and
transmission of light through the quantum wells for the
light-hole-to-electron absorption edge is also presented. 
\end{abstract}

\maketitle

\section{Introduction}
The gyromagnetic ratio or $g$-factor determines the splitting of spin
sublevels in the external magnetic field and it is one of the key
parameters describing energy spectrum of charge carriers in
semiconductors. It is well known that the electron and hole
$g$-factors in bulk semiconductors differ considerably from the free
electron $g$-factor value~\cite{PhysRev.114.90}. This difference is
caused by the spin-orbit interaction and $\bm k \cdot \bm
p$ mixing of the electron bands. In quantum well (QW) structures the size
quantization leads to an additional strong renormalization of the
$g$-factor as shown, e.g., in
Refs.~\cite{ivchenko_kiselev92,PhysRevB.75.245302} for the conduction band electrons 
(see also references cited in the book~\cite{ivchenko05a}) and in 
Refs.~\cite{efros1,kiselevmoiseev96_rus} for heavy holes in the valence band.
As for the light hole $g$-factor in QW structures, the
detailed calculations were carried out only for the in-plane components
$g_{xx}=g_{yy}$ (lateral or transversal Zeeman
effect)~\cite{kiselev_inplane}. The Zeeman effect on two-dimensional
light hole excitons in the magnetic field directed along the structure growth axis $z$ was studied in
Refs.~\cite{carmel,efros2,10.1063/1.2245213,PhysRevB.82.195317}.
However, no microscopic calculations of the longitudinal $g$-factor
component $g_{zz}$ were carried out. The present theoretical work concerns the effect of
quantum confinement on the longitudinal hole $g$-factor 
in the lowest light-hole subband $lh1$. It is shown below
that this $g$-factor component is determined to a great
extent by the proximity of the valence heavy-hole ($hh2$) and
light hole ($lh1$) subbands and their interface mixing.

Let us recall that, in zinc-blende lattice semiconductors, the Bloch
states at the 
valence band top transform according to four-dimensional spinor
representation $\Gamma_8$ of the T$_{d}$ symmetry point 
group. Under symmetry operations, the corresponding basis functions
transform similarly to 
the spinor spherical harmonics ${\cal Y}^{(l)}_{Jm}$ with the total
angular momentum  $J=3/2$  
and the orbital angular momentum $l=1$ ($m=-3/2,-1/2,1/2,3/2$). The
Zeeman interaction of a bulk hole with  
the magnetic field  ${\bm   B}$ is described by the
4$\times$4 matrix operator~\cite{birpikusbook}  
\begin{equation}
\mathcal H_0 = -2\mu_B[\varkappa (\bm{J}\cdot \bm{B}) + q(J_x^3 B_x+J_y^3
B_y+J_z^3 B_z)]\:.
\label{ham-0}
\end{equation} 
Here we use the coordinate frame $x||[100]$, $y||[010]$, $z||[001]$,
$\bm{J} = (J_x, J_y, J_z)$ is the vector composed of the  angular
momentum $J=3/2$ matrices, $\mu_B$ is Bohr magneton, $\varkappa$ and
$q$ are the band structure parameters. In what follows the small contribution
proportional to the constant $q$ and responsible for the anisotropic
Zeeman splitting is disregarded~\cite{Mar99}. The constant  $\varkappa$  
is related to the dimensionless Luttinger parameters $\gamma_i$
($i=1,2,3$)~\cite{PhysRev.102.1030,PhysRev.133.A542}.
For instance, in GaAs crystals,  $\varkappa$ = 1.2.   

A uniaxial strain of the bulk crystal results in the splitting of the
heavy- and light-hole $\Gamma$-point states, with angular momentum projections
$\pm 3/2$ and $\pm 1/2$ onto the deformation axis. In accordance with
Eq.~\eqref{ham-0} the Zeeman splittings of these states amount to  $-6\varkappa \mu_B B$  and
$-2\varkappa \mu_B B$, respectively. Hence, the light holes are described
by an effective $g$-factor of $g_{\rm eff} = - 2 \varkappa$. In
GaAs/AlGaAs QW structures grown along $[001]$ direction, the states of
the heavy ($hh$) and light 
($lh$) holes at the $\Gamma$-point, i.e., the states with zero lateral
wave vector, are quantized independently and two series of quantum-confined hole
states are formed: $hh \nu$ and $lh \nu$ ($\nu = 1,2...$). Here we will
demonstrate that the $g$-factor of light holes $lh1$ strongly differs
from the above 
value $- 2 \varkappa$, where $\varkappa$ is
averaged over the hole wave-function distribution between QW and
barrier layers.  The origin of the giant renormalization of  $g_{\rm
  eff}(lh1)$ is related to the fact that, in
GaAs/Al$_x$Ga$_{1-x}$As-type QWs, at the $\Gamma$ point the $lh1$ and
$hh2$ subbands  
are very close in energy. These subbands are intermixed owing to 
off-diagonal elements $H=\sqrt{3}\hbar^2\gamma_3
\hat{k}_z (k_x - \mathrm i k_y) /m_0$ and $H^*$ of the Luttinger
Hamiltonian~\cite{ivchenko05a}, where $m_0$ is the free electron mass,
$\gamma_3$ is one of the Luttinger parameters, $\hat{k}_z = - {\rm
  i} \partial/ \partial z$, and $k_x, k_y$ are the components of the
in-plane wave vector  ${\bm 
  k}_{\parallel}$. Indeed, in a magnetic field applied along the
$z$ axis, the vector ${\bm k}_{\parallel}$ acquires a
contribution  $- (e{\bm A}/c \hbar) $  that is proportional to the vector potential  ${\bm A}$ of
magnetic field ${\bm B}$. The allowance for this contribution results
in the mixing of the states $lh1$ and $hh2$. In the second order of the ${\bm
  k}\cdot {\bm p}$ perturbation theory we obtain for the
effective light hole $g$-factor:
\begin{equation} \label{g-factor-eff-1}
g_{\rm eff}(lh1) = \frac{E_{lh1,1/2}(B_z) - E_{lh1,-1/2}(B_z)}{\mu_B
  B_z} =
\end{equation}
\[
= -2\varkappa + 12 \frac{\hbar^2}{m_0} \frac{\left|\left\langle
      hh2|\gamma_3\hat{k}_{z}|lh1\right\rangle \right|^2}{E_{hh2} - E_{lh1}}\:, 
\]
where $|lh1\rangle$ and $|hh2\rangle$ are the hole wavefunctions
describing their size quantization along $z$-axis, $E_{lh1}$ and
$E_{hh2}$ are the size-quantization energies of the states $lh1$ and
$hh2$ in QWs at $\bm k_\parallel = 0$
(in the hole representation the energies $E_{lh1}, E_{hh2}$
are positive). It is worth to note that, in contrast to the studied geometry ${\bm
  B} \parallel z$, the magnetic field applied in the QW
plane can be described by the vector potential ${\bm A} =
(B_y z, - B_x z, 0)$ linear in $z$. In this case, the component
$H$ of Luttinger Hamiltonian is invariant under the mirror reflection
$z \to -z$. Therefore, in symmetric wells the states $lh1$ and $hh2$
are not mixed by the in-plane magnetic field and the 
resonant contribution to the lateral $g$-factor is absent. 

In the limiting case of the infinite barries one has
\[
E_{lh1} = \frac{\hbar^2\pi^2}{2m_0a^2}(\gamma_1 + 2\gamma_2)\:,\quad \:E_{hh2} = \frac{4\hbar^2\pi^2}{2m_0a^2}(\gamma_1 - 2\gamma_2)\:,
\]
where $a$ is the well width, and the matrix element in
Eq.~(\ref{g-factor-eff-1}) is given by 
$$\langle
hh2|\gamma_3\hat{k}_{z}|lh1\rangle =\frac{8{\rm i}\gamma_3}{3a}\:.$$
An analogous expression where $\varkappa$ is replaced by
$3\varkappa$ and the second term is taken with the reversed sign describes the
effective $g$-factor of the heavy hole in the subband
$hh2$. The formula~(\ref{g-factor-eff-1}) can be easily extended for
calculation of the $g$-factor in the heavy-hole $hh1$ subband in hybrid
deformed QWs studied recently in
Ref.~\cite{PhysRevB.83.165450} where the states  $lh1$ and $hh1$ can
be resonant. In this case, the matrix element $\langle
hh1|\gamma_3 \hat{k}_{z}|lh1\rangle$ is non-zero due to the structure
asymmetry as well as to an external electric field applied along the growth axis.

Following Ref.~\cite{efros2} we introduce an effective mass of
the light hole $lh1$, which, in the resonant approximation of the $\bm
k \cdot \bm p$ perturbation theory, can be presented as
\begin{equation} \label{mass}
\frac{m_0}{m_{lh1}} = \gamma_1 - \gamma_2 + 6 \frac{\hbar^2}{m_0} \frac{\left|\left\langle
      hh2|\gamma_3\hat{k}_{z}|lh1\right\rangle \right|^2}{E_{lh1} - E_{hh2}}\:.
\end{equation}
Comparing Eqs.~(\ref{g-factor-eff-1}) and (\ref{mass}), we obtain
the relation between the $g$-factor and the effective mass
\[
g_{\rm eff}(lh1) = 2 \left( - \varkappa + \gamma_1 - \gamma_2 - \frac{m_0}{m_{lh1}}  \right)\:.
\]
This expression differs from the analogous relation, see Eq. (6) in
Ref.~\cite{efros2}, by the sign of $\varkappa$.

It follows from Eq.~\eqref{g-factor-eff-1} that, for the infinite
barriers, the dependence of the $g$-factor on the QW width vanishes
and is  given by
\begin{equation}
g_{\rm eff}{(lh1)} = -2\varkappa-\frac{512}{3\pi^2}
\frac{\gamma_3^2}{10\gamma_2 - 3\gamma_1} \approx -26\:.
\label{g:inf}
\end{equation}
The estimate is valid in the spherical approximation where
$\gamma_2 = \gamma_3 = \bar \gamma$ (see Table~\ref{table:parameters}
for the parameter values).
Thus, indeed, the proximity of the ground light and excited heavy hole
subbands results in a giant enhancement of the light-hole
Zeeman splitting. A similar enhancement effect for the $\bm k$-linear
spin-dependent terms in the light hole Hamiltonian was predicted
by Rashba and Sherman~\cite{Rashba1988175}.

It is instructive to analyze the effect of all heavy hole subbands on
the light-hole $g$-factor enhancement. In the limit of infinite barriers the
summation over all even subbands can be carried out analytically with the result
\begin{equation}
g_{\rm eff}(lh1) =
\label{gfactor_cot}
\end{equation}   
\[
-2\varkappa-12\frac{\gamma_3^2(\nu+1)}{\gamma_1(\nu-1)}\left[ 1 +
  \frac{4\sqrt{\nu}}{\nu-1}\frac{\cot\frac{\pi}{2}\sqrt{\nu}}{\pi}
\right] \approx -24.7,
\]
where $\nu =
(\gamma_1+2\gamma_2)/(\gamma_1-2\gamma_2)$.

To conclude the introduction, we would like to point out that in real systems the absolute
values and even the sign of effective $g$-factor may be extremely sensitive
to the barrier height and QW width. Moreover, for large
magnitudes of  $|g_{\rm
eff}|$, the Zeeman splitting becomes
non-linear function of the field, even in moderate magnetic fields. These effects are also addressed below.

\section{Allowance for the interface heavy-light-hole mixing}
\label{sec:gfactor}

Low C$_{2v}$ point symmetry of the ideal $(001)$ interface allows for
the light-heavy hole mixing even at ${\bm k}_\parallel
=0$~\cite{aleiner92,ivchenko96}. In the well with symmetric interfaces
the corresponding contribution to the effective Hamiltonian can be
written as~\cite{aleiner92,ivchenko96,toropov:035302,ivchenko05a} 
\begin{equation}
\mathcal H_{l-h} = \pm
\frac{t_{l\mbox{-}h}\hbar^2}{\sqrt{3}m_0a_0}\left\{J_xJ_y\right\}\delta(z-z_i)\:, 
\label{mixing}
\end{equation}
where $a_0$ is the lattice constant, $t_{l-h}$ is the
dimensionless mixing parameter of the order of unity,
$\left\{J_xJ_y\right\} = (J_xJ_y + 
J_yJ_x)/2$ is the symmetrized product of the angular momentum
operators, $z_i$ are the coordinates of
interfaces. Hereafter we consider only the $lh1$ and $hh2$
hole states, assuming that the energy gap between them
as well as their Zeeman splittings much smaller than the energy
distance to other quantized energy states. 

\begin{figure}[tb]
\includegraphics[width=0.48\textwidth]{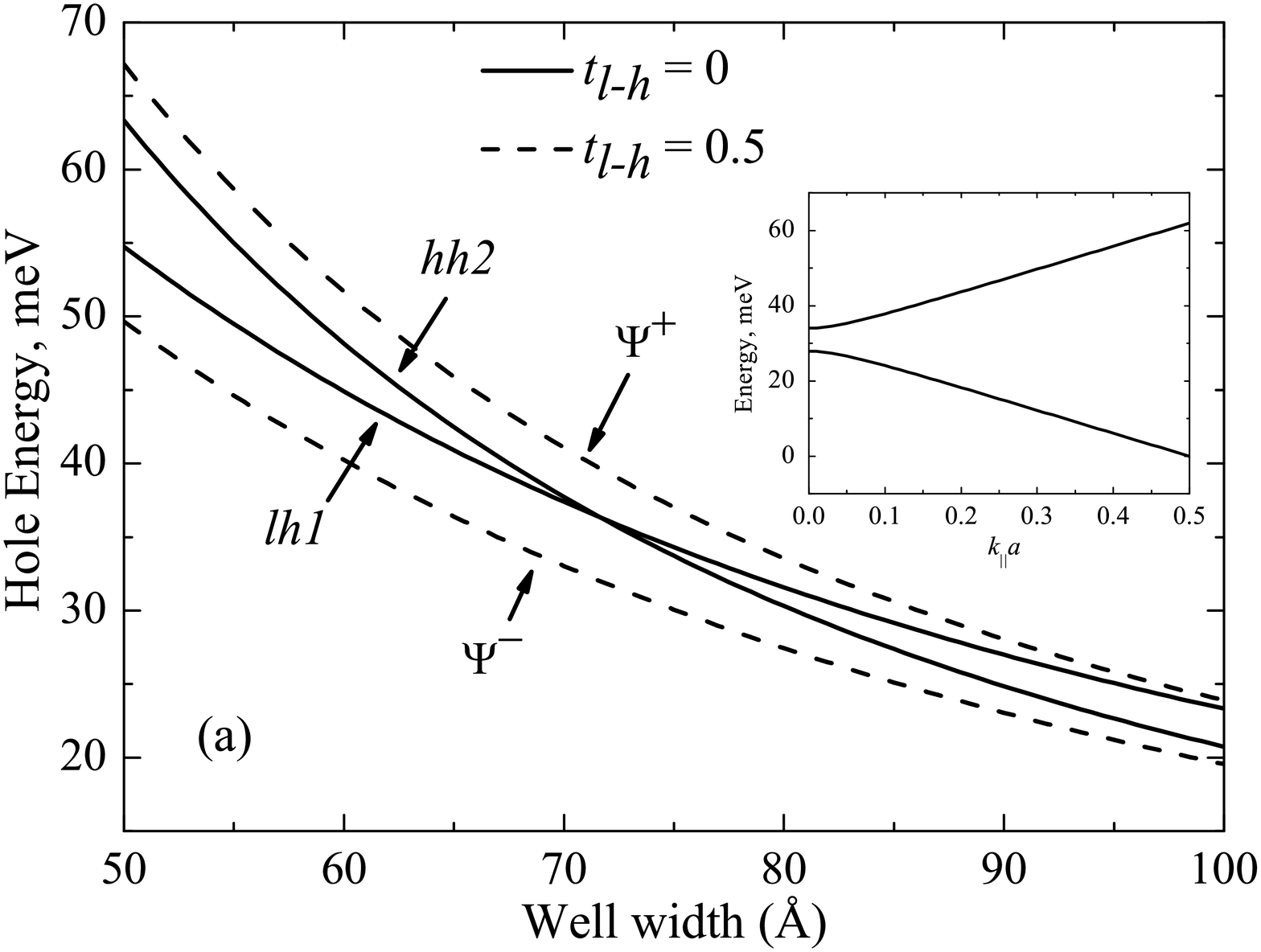}
\includegraphics[width=0.48\textwidth]{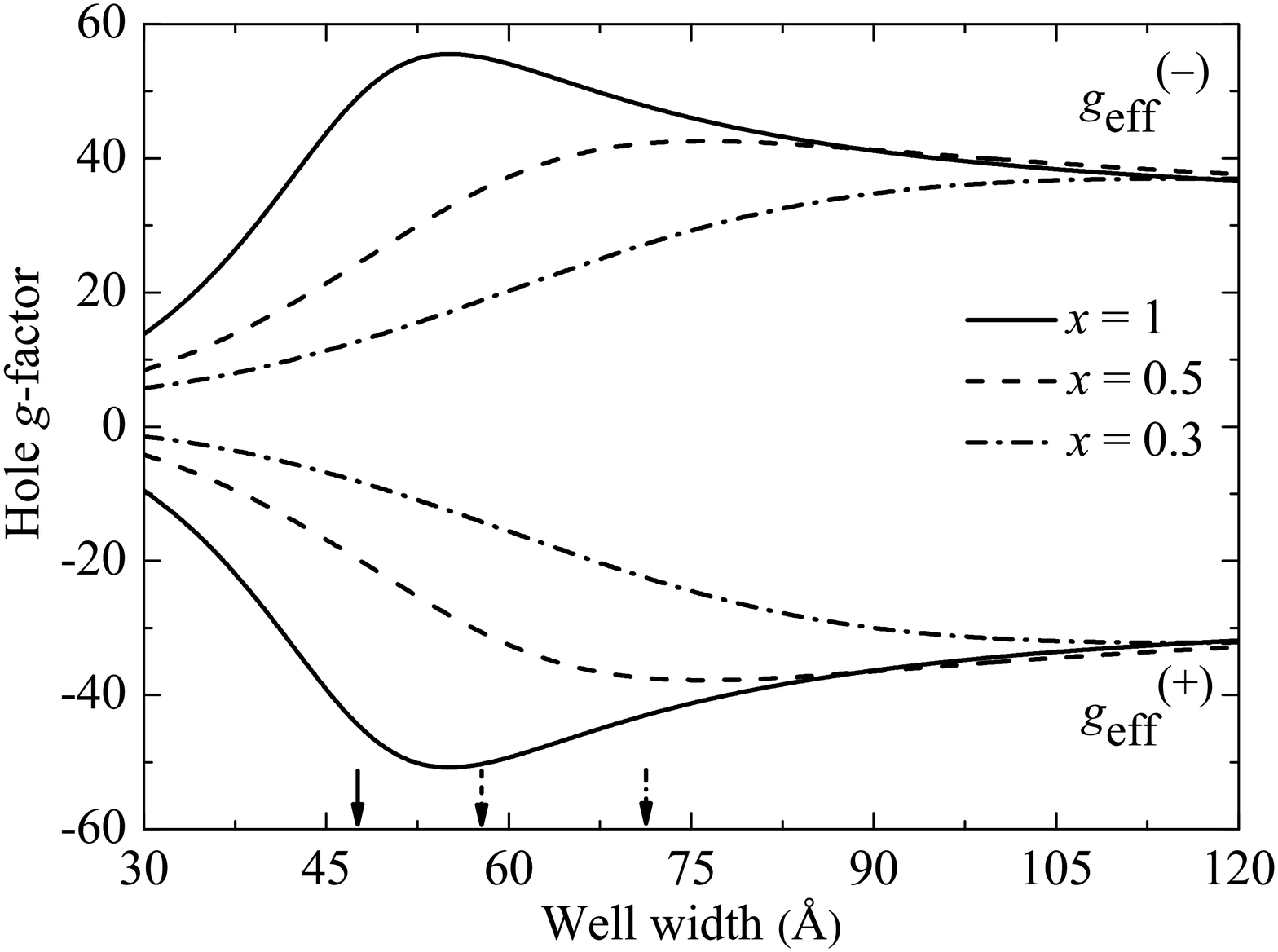}
\caption{(a) Heavy ($hh2$) and light ($lh1$) hole energy as a function
  of the GaAs/Al$_{0.3}$Ga$_{0.7}$As QW width calculated for the two
  values of the interface mixing parameter: $t_{l\mbox{-}h}$ = 0
  (solid lines) and  $t_{l\mbox{-}h}$ = 0.5 (dash lines). Inset shows
  energy dispersion of the hole subbands for the well width $a = 80$
  \AA. (b) The $g$-factors $g_{\rm eff}^{(\pm)}$ of the split energy
  states $E^+$ and $E^-$ as functions of QW width for
    three compositions $x$ and the mixing parameter $t_{l\mbox{-}h}$
  = 0.5. Vertical arrows indicate values of the
  anticrossing points $a_{\rm cr}$ for each $x$.} 
\label{fig:levels}
\end{figure}

Figure~\ref{fig:levels}(а) shows the results of calculation of
the heavy ($hh2$) and light ($lh1$) hole energy dependence on the
GaAs/\ Al$_{0.3}$Ga$_{0.7}$As QW width. The parameters used in
calculation are summarized in
Table~\ref{table:parameters}. In the applied spherical model, the values of
$\gamma_2$ and $\gamma_3$ are replaced by the average $\bar \gamma = (2 \gamma_2 + 3 \gamma_3)/5$. Solid lines
represent the energies calculated disregarding the
interface mixing, i.e., at $t_{l\mbox{-}h}=0$ in which case the $lh1$ and
$hh2$ energy branches cross each other at the well width $a_{\rm cr}
\approx 70$~\AA. Dashed lines show the calculation performed
including the interface mixing effect (with
$t_{l\mbox{-}h}=0.5$) which results in anticrossing of the $lh1$ and
$hh2$ subbands and formation of the hybrid states  
\begin{eqnarray} \label{wave-functions}
&&\Psi^+_{\pm 1/2} = C_l^+ \left| lh1, \pm 1/2 \right\rangle \pm C_h^{+} \left| hh2, \mp 3/2 \right\rangle \:, \\
&&\Psi^-_{\pm 1/2} = C_l^{-} \left| lh1, \pm 1/2 \right\rangle \pm C_h^{-} \left| hh2, \mp 3/2 \right\rangle \:. \nonumber
\end{eqnarray} 
Here the complex coefficients $C_h^{-}, C_l^{-}$ and $C_h^{+},
C_l^{+}$ are interrelated by the ortogonality 
condition, in particular, $|C^+_l|^2 = |C^-_h|^2$ and
$|C^-_l|^2 = |C^+_h|^2$; the superscript ``+'' or $``-$'' in
$\Psi^{\pm}_{\pm 1/2}$ denotes the upper and lower states
with energies $E^+ > E^-$, respectively; the subscript $\pm 1/2$
enumerates degenerate states, it coincides with the projection of the
angular momentum of the light-hole state admixed to 
$\Psi^+_{\pm 1/2}$ or $\Psi^-_{\pm 1/2}$. The energies $E^{\pm}$ of mixed
states and the coefficients in (\ref{wave-functions}) are
determined in accordance with the procedure developed
in~\cite{toropov:035302}.  

At the $\Gamma$ point, the $hh2 \to e1$ optical transitions are
forbidden and $lh1 \to e1$ transitions are allowed. It is therefore
convenient to define $g$-factors of the $\Psi^+_{\pm 1/2}$ and
$\Psi^-_{\pm 1/2}$ states as follows    
\begin{equation} \label{mixing2}
g_{\rm eff}^{\pm} = \frac{E^{\pm}_{1/2}(B_z) - E^{\pm}_{-1/2}(B_z)}{\mu_B B_z}\:.
\end{equation}
The $g$-factors defined in such a way are given by
\begin{eqnarray} \label{geff+-}
g_{\rm eff}^{(+)} = - 2 \varkappa (|C^+_l|^2 - 3 |C^+_h|^2) - \Delta g_{\rm eff} \:,\\
g_{\rm eff}^{(-)} = - 2 \varkappa (|C^-_l|^2 - 3 |C^-_h|^2) + \Delta g_{\rm eff} \:, \nonumber
\end{eqnarray}
where $\pm \Delta g_{\rm eff}$ are the contributions to the Zeeman effect due to the magnetic field induced mixing of the split states $\Psi^+_{\pm 1/2}$ and $\Psi^-_{\pm 1/2}$ with
\begin{equation} \label{Deltag}
\Delta g_{\rm eff} = 12 \frac{\hbar^2}{m_0} \frac{\left|\left\langle
      hh2|\gamma_3\hat{k}_{z}|lh1\right\rangle \right|^2}{E^+ -
  E^-}\:.
\end{equation}
This formula differs from the corresponding contribution in
Eq.~(\ref{g-factor-eff-1}) by the replacement of the
denominator $E_{hh2} -  E_{lh1}$ 
by $E^+ -  E^-$, if $E_{hh2} < E_{lh1}$, and by $-(E^+-E^-)$,
if $E_{hh2}<E_{lh1}$. Hence, far from the anticrossing
point $a_{\rm cr}$, Eq.~\eqref{geff+-} for $g_{\rm eff}^{(-)}$ at $a <
a_{\rm cr}$ and that for $g_{\rm eff}^{(+)}$  
at $a > a_{\rm cr}$ transform into
Eq.~\eqref{g-factor-eff-1}. We note that the
allowance for the penetration of hole wavefunctions into
barriers results in a more complicated expression for the first term in the 
right-hand side of Eq.~(\ref{geff+-}) representing the ``bulk'' contribution to the
Zeeman effect. The estimations however show that this difference is negligible. The negative sign
in front of $3 |C^+_h|^2$ and $3 |C^-_h|^2$ takes into account
the state $| lh1,\pm 1/2 \rangle$ mixes with $| hh2,\mp 3/2
\rangle$. The interface mixing leads to an existence of gap between
the energies $E^+$ and $E^-$  
independently of the width of the structure and finite $g$-factor
values, even within the second order of 
perturbation theory. Therefore, the allowance for the
interface mixing effects in the hybrid deformed
structures~\cite{PhysRevB.83.165450} should lead to finite values of  
$g$-factors of the heavy and light holes for any value of the electric
field.       

\begin{table}
\caption{
Parameters used in the calculation of light hole $g$-factor. In case
of the alloys, the Luttinger parameters and the $\varkappa$
constant are obtained by a linear interpolation of the
corresponding values for GaAs and AlAs~\cite{vurgaftman02}, and the parabolic interpolation is applied to determine
the valence band offsets.} 
 \begin{tabular}{p{0.2\linewidth} p{0.1\linewidth} p{0.1\linewidth} p{0.1\linewidth} p{0.1\linewidth} c c}
\hline
\hline
Material  & $\gamma_1$ & $\gamma_2$ & $\gamma_3$ & $\bar{\gamma}$ & $\Delta E_v$, meV & $\varkappa$ \\
\hline
GaAs & 6.98 & 2.06 & 2.93 & 2.58 & -- & 1.2\\
Al$_{0.3}$Ga$_{0.7}$As  & 6.01 & 1.69 & 2.48 & 2.16 & 140 & 0.87\\
%Al$_{0.36}$Ga$_{0.64}$As  & 5.82 & 1.61 & 2.38 & 2.07 & 140 & 0.87\\
Al$_{0.5}$Ga$_{0.5}$As  & 5.37 & 1.44 & 2.17 & 1.88 & 255 & 0.66\\
AlAs & 3.76 & 0.82 & 1.42 & 1.18 & 600 & 0.12\\
In$_{0.53}$Ga$_{0.47}$As  & 13.7 & 5.42 & 6.31 & 5.95 & 354 & 4.63\\
InP & 5& 1.6 & 2 & 1.84 & -- & 0.97 \\
\hline
\hline
\end{tabular}
\label{table:parameters}
\end{table}

Figure~\ref{fig:levels}(b) demonstrates the calculated
values of $g_{\rm eff}^{(+)}$ and $g_{\rm eff}^{(-)}$ as functions of
GaAs/Al$_x$Ga$_{1-x}$As QW width for different contents
$x$ of Al in the barriers. It is clear that the
magnetic-field induced light-heavy hole mixing results in a strong
increase of the absolute value of $g$-factor as compared to the bulk
material. One can also see that the absolute values of $g$-factor
increase with increasing $x$ which is a consequence of the rising barrier height. In wide
wells $g_{\rm eff}^{(+)}$ tends to its asymptotic
value given by Eq.~\eqref{g:inf} because in this limiting case the wavefunction
penetration in barriers can be neglected.

The absolute value of $g$-factor approaches its maximum in
the vicinity of the anticrossing point of the $lh1$ and $hh2$ subbands.
If we declared the mixed states (\ref{wave-functions}) to be 
the heavy-hole states as soon as $|C_l|^2 < |C_h|^2$ and light-hole
states if $|C_l|^2 > |C_h|^2$, 
then at the anticrossing point the $g$-factors of the so-defined hole
states would exhibit 
an abrupt discontinuity and sign reversal.     

Since the $g$-factors (\ref{geff+-}) are governed by the difference of
the unperturbed $hh2$ and $lh1$ energies which is extremely sensitive
to the Luttinger parameters, even small variations of these  
parameters can considerably alter the dependence of $g_{\rm eff}^{\pm}$ 
on the well width. 
\section{Resonant spectra of the optical transmission and reflection} 
Despite the development of the numerous techniques of the
optical spectroscopy, experimental data on the
$g$-factors of light holes in QWs are scarce. In
Refs.~\cite{PhysRevB.37.4171,PhysRevB.48.1955} the measured values of 
light hole $g$-factors are close to unity. More detailed investigation
of the $g$-factors of light and heavy holes is performed in
Refs.~\cite{efros2,10.1063/1.2245213} by differential 
magnetoabsorption and magnetotransmission techniques. For the
theoretical analysis of these effects let us recall that, for a
single-QW structure with one excitonic resonance taken into account, the
light reflection ($R$) and transmission ($T$) coefficients can be
written as   
\begin{eqnarray} \label{single}
R &=& \left\vert r_{01} + \frac{{\rm e}^{2 {\rm i} \theta} t_{01} t_{10}r_{QW}}{ 1 - r_{10} r_{QW} {\rm e}^{2 {\rm i} \theta}} \right\vert^2 \approx R_0 [1 + S_0 f(x, \Phi)]\:,\\
T &=& n \left\vert \frac{t_{01} {\rm e}^{{\rm i} \theta} ( 1 + r_{QW})}{ 1 - r_{10} r_{QW} {\rm e}^{2 {\rm i} \theta}} \right\vert^2 \approx T_0 \left[ 1 - Q_0 h(x) \right]\:, \nonumber
\end{eqnarray}
where
$$
R_0 = r_{01}^2\:,\:T_0 = n |t_{01}|^2\:,\:$$
$$ r_{01} = - r_{10} = - \frac{n-1}{n+1}\:,\:t_{10} = n t_{01} = \frac{2 n}{n+1}\:,
$$
and $r_{QW}$ is the amplitude reflection coefficient from the
QW, $\theta$ is the phase shift due to the light
propagation over the distance between the external
boundary ``vacuum $-$ cap layer'' and the QW center,
$$
S_0 = \frac{8 n}{n^2 - 1} \frac{\Gamma_0}{\Gamma}\:,\:Q_0 = \frac{2\Gamma_0}{\Gamma}\:,
$$
$$
f(x, \Phi) = \frac{\sin{\Phi} + x \cos{\Phi}}{x^2 + 1}\:,\:h(x) = \frac{1}{x^2 + 1}\:,$$
$$\:
\:x = \frac{\omega - \omega_0}{\Gamma}\:,\:
\Phi = 2 \theta + \frac{\pi}{2}\:. 
$$
Other notations are common: $n$ is the refraction index (we neglect the
difference of its values in the well and the barriers),
$\omega_0, \Gamma_0$ and $\Gamma$ are the resonant frequency,
radiative and nonradiative excitonic dampings,
respectively. Note, that the  parameter $\Gamma$
describes, in fact, the inhomogeneous
broadening. Hereafter we assume that $\Gamma_0
\ll \Gamma$ and neglect terms quadratic in $r_{QW}$. The transmission
coefficient $T$ is defined as 
the ratio of the incident radiative flux and the flux escaping through
the cap layer and the QW to the semi-infinite
barrier.          

\begin{figure}[hptb]
\includegraphics[width=\linewidth]{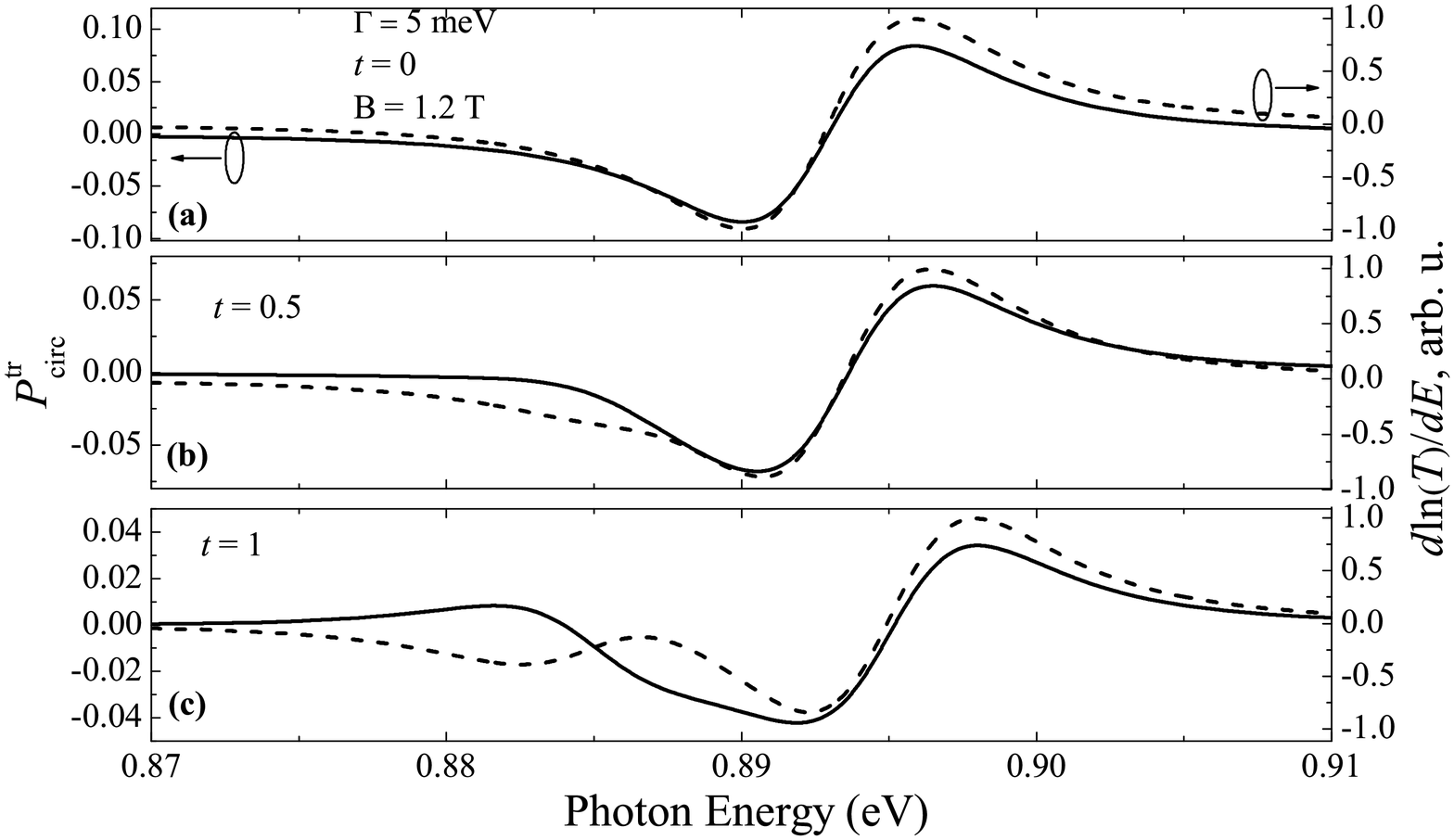}\\
\includegraphics[width=\linewidth]{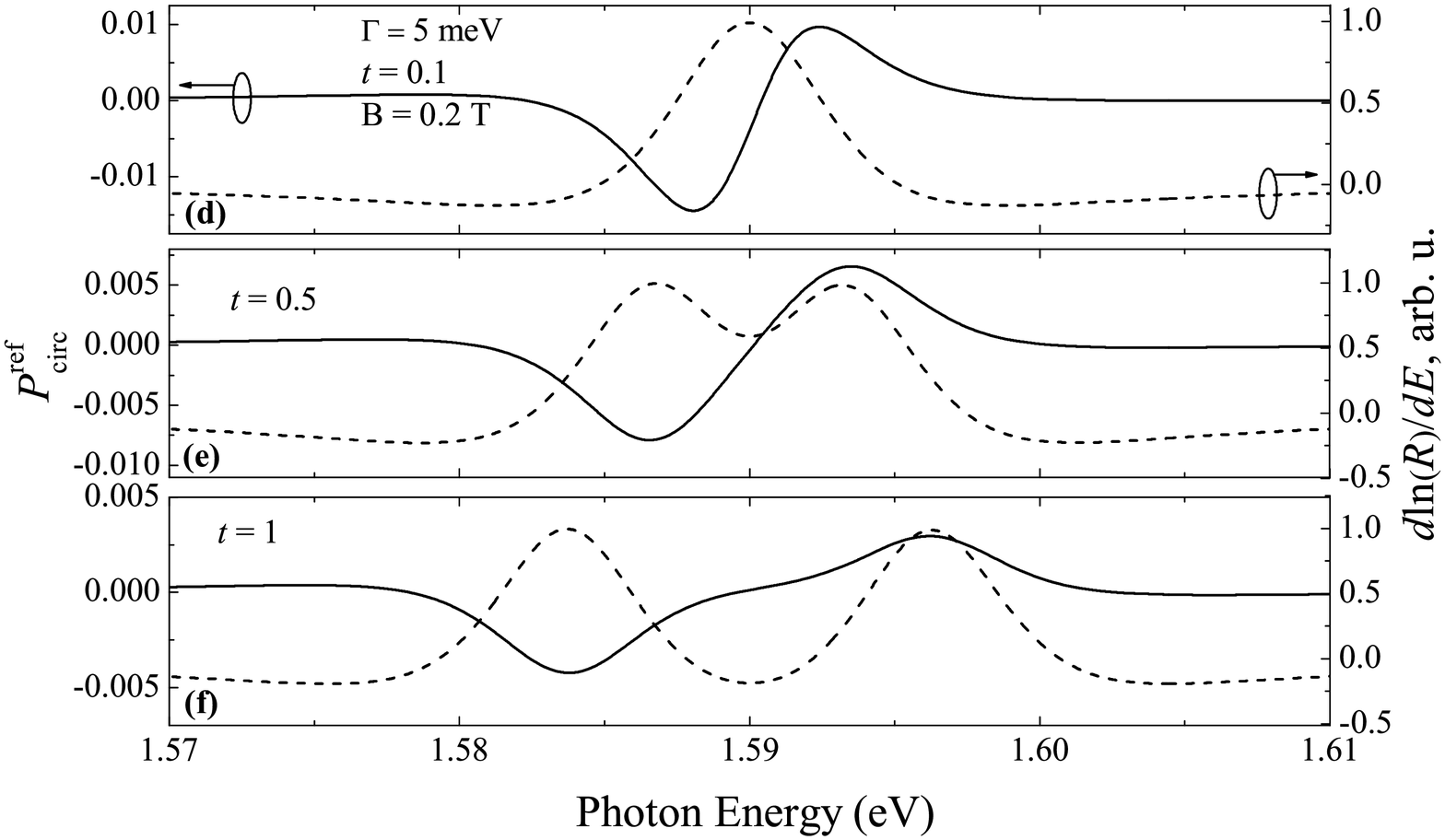}
\caption{(a) -- (c) Transmission spectra  $P_{\rm circ}^{\rm tr}$
  (solid curves) and $d\ln{T}/d(2E)$ (dashed curves) through the structure
  with 100~\AA~In$_{0.53}$Ga$_{0.47}$As/GaAs QW. The panels
  are computed for different strengths of heavy-light hole mixing
  at interfaces: $t_{l-h} = 0$, $0.5$ and $1$,
    respectively. The nonradiative broadening $\hbar \Gamma=5$~meV. (d) -- (f)
  Reflection spectra $P_{\rm circ}^{\rm ref}$ (solid curves) and
  $d\ln{R}/dE$ (dashed curves) from the structure with
  80~\AA~GaAs/Al$_{0.36}$Ga$_{0.64}$As QW. The panels are
  computed for $t_{l-h} = 0.1$, $0.5$ and $1$. The nonradiative
  broadening $\hbar \Gamma=5$~meV, 
 the phase $\Phi$ is a multiple of $2 \pi$.}
\label{fig:spectra}
\end{figure} 
In Refs.~\cite{efros2,10.1063/1.2245213} the light-hole $g$-factor was
experimentally determined  
from measurements of the differential spectra of circular magnetotransmission
and magnetoreflection defined as follows
\begin{equation} \label{Pcirc}
P_{\rm circ}^{\rm tr} = \frac{T_{\sigma_+} - T_{\sigma_-}}{T_{\sigma_+} + T_{\sigma_-}}\:,\:P_{\rm circ}^{\rm ref} = \frac{R_{\sigma_+} - R_{\sigma_-}}{R}\:,
\end{equation}
where $T_{\sigma_{\pm}}$ and $R_{\sigma_{\pm}}$ are the
spectrally dependent intensity coefficients of transmission
and reflection of the circularly polarized light
$\sigma_{\pm}$, respectively, $R = (R_{\sigma_+} +
R_{\sigma_-})/2$. Replacing the resonant frequency $\omega_0$ in
Eq.~(\ref{single}) by the frequencies $\omega_{0, \pm} = \omega_0 \pm
\Delta E/(2 \hbar)$, where $\Delta E$ is the Zeeman splitting of the
sublevels $lh1, 1/2$ and $lh1, - 1/2$, we obtain the relationship
between the differential spectra and the value of $\Delta E$ for the
excitonic resonance $lh1$ 
\begin{equation} \label{experP}
P_{\rm circ}^{\rm tr} = - \frac{\Delta E}{2} \frac{d \ln{T(E)}}{d E}
\approx - \frac{\Delta E}{2 T_0}~ \frac{d T(E)}{d E} \:,
\end{equation}
\[
P_{\rm
  circ}^{\rm ref} = - \frac{\Delta E}{2} \frac{d \ln{R(E)}}{d E}
\approx - \frac{\Delta E}{R_0}~ \frac{d R(E)}{d E}\:, 
\]

where $E = \hbar \omega$, $T_0$ and $R_0$ are the transmission and
reflection coefficients in the absence of magnetic field. We
stress that these formulae
are based on the assumption of a single excitonic level in the region
of excitonic resonance. However, in the region of anticrossing between
$e1$-$lh1$ and $e1$-$hh2$ excitons, caused by interface mixing of
heavy and light holes, optical spectra are determined by two close
resonances $e1$-$\Psi^+$ and $e1$-$\Psi^-$ and, strictly speaking, 
Eq.~(\ref{experP}) is invalid. For this reason, we have derived 
an expression for the differential reflection spectra for a pair 
of closely-lying excitonic levels and present. The result for two resonances
reads
\begin{equation}
R = R_0 \{ 1 + S_0 [ |C^+_l|^2 f(x_+, \Phi) + |C^-_l|^2 f(x_-, \Phi)] \}\:,
\end{equation}
where $x_{\pm} = (\omega - \omega_{0,\pm})/\Gamma$. Here, for simplicity, we neglect the difference between 
the reduced masses of $e1$-$\Psi^+$ and $e1$-$\Psi^-$ excitons. Nonradiative
decays $\Gamma_+$ and $\Gamma_-$ are considered to be equal. In the
presence of magnetic field one has  
\[
R_{\sigma_{\pm}} = R_0 \{ 1 + S_0 [ |C^+_l|^2 f(x_+ \mp \delta_+,
\Phi) + |C^-_l|^2 f(x_- \mp \delta_-, \Phi)] \}\:, 
\]
\[
\delta_{\pm} = \frac{g_{\rm eff}^{(\pm)} \mu_B B_z}{2 \hbar \Gamma}\:.
\]
For the differential circular reflection one obtains
\begin{equation} \label{two_res_ref}
P_{\rm circ}^{\rm ref} = \frac{R_{\sigma_+} - R_{\sigma_-}}{R_0} = 
\end{equation} 
\[ 
= -\frac{\mu_B B_z}{\hbar  
\Gamma}
S_0 [ |C^+_l|^2 g_{\rm eff}^{(+)}  f'(x_+, \Phi) +  |C^-_l|^2 g_{\rm
  eff}^{(-)}  f'(x_-, \Phi)]\:, 
\]
\[
f'(x, \Phi) = \frac{\partial f(x, \Phi)}{\partial x} = - \frac{(x^2 -
  1) \cos{\Phi} + 2 x \sin{\Phi}}{(x^2 + 1)^2}\:. 
\]
Let us also present the expression for the relative
differential reflection in the absence of magnetic field 
\begin{equation} \label{dif_ref}
\frac{1}{R_0}~ \frac{d R(\hbar \omega)}{d(\hbar \omega)} = \frac{S_0}{\hbar 
\Gamma} [ |C^+_l|^2  f'(x_+, \Phi) +  |C^-_l|^2   f'(x_-, \Phi)]\:.
\end{equation}
One can see that in the presence of two close resonances the ratio of $P_{\rm
  circ}^{\rm ref}$ and $d \ln{R}/dR$ is not a constant but rather it
is a function of frequency and may even change its sign within the
linewidth.     

For the transmission spectra, the expressions analogous
  to Eqs.~(\ref{two_res_ref})~and~(\ref{dif_ref})  
have the form
\begin{equation} \label{two_res_tr}
P_{\rm circ}^{\rm tr} = \frac{T_{\sigma_+} - T_{\sigma_-}}{2 T_0} =
\end{equation}
\[
=\frac{\mu_B B_z}{2 \hbar  
\Gamma} Q_0 [ |C^+_l|^2 g_{\rm eff}^{(+)}  h'(x_+) +  |C^-_l|^2 g_{\rm
eff}^{(-)}  h'(x_-)]\:,
\]
\[ 
\frac{1}{T_0}~ \frac{d T(\hbar \omega)}{d(\hbar \omega)} = - \frac{Q_0}{\hbar 
\Gamma} [ |C^+_l|^2  h'(x_+) +  |C^-_l|^2   h'(x_-)]\:,
\]
\[\: h'(x) = -
\frac{2x}{(x^2 + 1)^2} \:. 
\]

Figure~\ref{fig:spectra} presents spectra of the differential
circular transmission $P_{\rm circ}^{\rm tr}$, panels (a)--(c), and
reflection $P_{\rm circ}^{\rm ref}$, panels (d)--(f), calculated
following Eqs.~(\ref{two_res_tr})~and~(\ref{two_res_ref}),
respectively. The parameters of calculations are given
  in the figure caption. The reflectivity is calculated for the system
studied in the work~\cite{10.1063/1.2245213} while the transmissivity 
is calculated for the system studied in the work~\cite{efros2}.  For comparison, the dashed lines
show the spectra of $d\ln{R}/dE$ and $d\ln{T}/dE$ calculated from
Eqs.~(\ref{dif_ref}) and formula in Eq.~(\ref{two_res_tr}). Let us emphasize that in the single-resonance model
the solid and dashed curves would be geometrically
  similar. In the InGaAs/GaAs 
structure~\cite{efros2}, the heavy-hole $hh2$
and light-hole $lh1$ states are rather distant in energy. Therefore, for
moderate values of the interface mixing parameter      
$t_{l-h} \leqslant 0.5$, panels (a) and (b), the difference in
behaviour of $P^{\rm tr}_{\rm circ}$ and $d\ln{T}/d(2E)$ is quite
small. If the interface mixing is significant, then as one can see in
Fig.~\ref{fig:spectra}(с), the spectra of $P^{\rm tr}_{\rm circ}$ and
$d\ln{T}/d(2E)$ calculated for $t_{l-h}=1$ have different qualitative
behaviour.      

A particularly interesting situation is realized if the
$lh1$ and $hh2$ states are close in energy. This case is
illustrated in Fig.~\ref{fig:zeeman-num}(d)--(f) representing
the differential reflection spectra $P_{\rm circ}^{\rm ref}$
calculated for the structure with a 80~\AA~ GaAs/AlGaAs quantum
well. Even at small strength of the interface mixing, $t_{l-h}=0.1$, see panel
(d), the spectral behaviours of $P_{\rm circ}^{\rm ref}$ and $d\ln{R}/dE$
are completely different. For example, let us consider the system tuned to the resonance between the bare
$hh2$ and $lh1$ hole states so that $E_{hh2} = E_{lh1}$ and $|C_l^+|^2 =
|C_l^-|^2$, $g_{\rm eff}^{(+)} \approx - g_{\rm eff}^{(-)}$. Moreover, let the system satisfy the condition 
$\omega_+ - \omega_- \ll \Gamma$. It follows then from Eq.~(\ref{two_res_ref})
that in this case the $P_{\rm circ}^{\rm ref}$ spectrum is in fact described by the second
derivative $f''(x, \Phi)$  whereas $d\ln{R}/dE \propto f'(x,\Phi)$. 

Our estimations of the light hole $g$-factors noticeably
exceed those extracted from the experiments \cite{efros2, 10.1063/1.2245213}. A detailed 
comparison between the developed theory and the existing experimental data
is out of the scope of the present paper because such a fitting requires
the inclusion of many free parameters and the better accuracy of
measurements. For a consistent description 
of experiments it is first of all necessary to determine the exact
energy positions of $lh1$ and $hh2$ states. The $lh1$-$hh2$ spacing is
very sensitive to the well width, barrier height and
Luttinger parameters which can lead to considerable variations of
$g_{\rm eff}$ values.

\section{Zeeman splitting at high magnetic fields}
\label{sec:zeeman}

Giant values of the $g$-factor obtained in the previous section by means of
the perturbation theory indicate that the Zeeman splitting of the
light-hole spin sublevels can 
deviate from the linear dependence already at moderate magnetic
fields. In this section we 
calculate the spin splitting beyond the linear approximation used in
the derivation of 
Eqs.~(\ref{g-factor-eff-1}) and (\ref{geff+-}) but again assume 
the energy gap between $lh1$ and $hh2$ states and the hole spin
splitting to be small as compared to the energy 
distances to other levels.

Neglecting the interface mixing of the hole states ($t_{l\mbox{-}h}=0$), the pair of states $|hh2; 3/2\rangle$,
$|lh1;1/2\rangle$ are uncoupled from the pair $|hh2; -3/2\rangle$,
$|lh1;-1/2\rangle$. In an external magnetic field the effective
Hamiltonian that describes the pair states is analogous
to the 2$\times$2 Hamiltonian of an electron in a two-dimensional
system with the spin-orbit splitting linear in the wavevector 
or to the Hamiltonian of ``massive'' Dirac
Fermions~\cite{bychkov84,RevModPhys.82.3045}. Decomposing the
hole wavefunction over the eigen functions of a charged
  particle in the magnetic
field~\cite{rashba64,PhysRevLett.96.086805,Averkiev2005543}, one
obtains for the Zeeman splitting   
\begin{equation}
\label{zeeman:t=0}
\Delta E_Z = \frac{\tilde{E}}{2} - \sqrt{\frac{\tilde{E}^2}{4} + 6\frac{\hbar^2}{m_0}\left|\left\langle
      hh2|\gamma_3\hat{k}_{z}|lh1\right\rangle \right|^2 \hbar \omega_c} {\color{red}\:.}
\end{equation}
Here $\omega_c = |e|B/m_0c$ is the cyclotron frequency of
a free electron in the magnetic field, and the $\tilde{E} = E_{lh1} -
E_{hh2} -(\gamma_1 + 2\gamma_2)\hbar 
\omega_c$. Equation~\eqref{zeeman:t=0} is derived assuming that
$E_{lh1}>E_{hh2}$, otherwise one should reverse the sign before the square root in Eq.~\eqref{zeeman:t=0}.   

\begin{figure}[hptb]
\includegraphics[width=\linewidth]{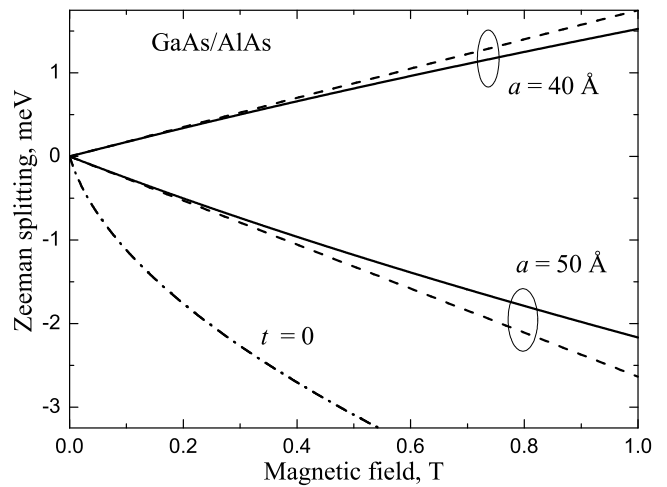}
\caption{Zeeman splitting of the light-hole subband $lh1$ in GaAs/AlAs QWs
  for two different widths: $a$ = 40~\AA~and $a$ = 50~\AA. Solid lines
  represent numerical calculations while dashed lines represent the
  linear-in-magnetic-field approximation following
  Eq.~(\ref{g-factor-eff-1}). The interface hole mixing
  parameter $t_{l\mbox{-}h} = 0.5$, the size quantized energies at
  zero field are $E_{lh1} = 111$~meV, $E_{hh2} = 119$~meV at $a =
  40$~\AA~and $E_{lh1} = 82.9$~meV, $E_{hh2} = 81.8$~meV at $a =
  50$~\AA. For comparison, dash-dot line shows the Zeeman splitting calculated from Eq.~\eqref{zeeman:t=0}
  in the absence of interface mixing for the particular well width
  $a=50$~\AA~ at which the $lh1$ and $hh2$ hole subbands almost touch
  each other at the $\Gamma$ point.}  
\label{fig:zeeman-num}
\end{figure}

It follows from Eq.~\eqref{zeeman:t=0} that the spin splitting $\Delta E_Z$
is a sublinear function of the field. Of course, in the
limit of weak fields where $|g_{\rm eff}| \mu_B B \ll |E_{lh1} -
E_{hh2}|$, the splitting is proportional to $B$ and the light-hole $g$-factor (\ref{zeeman:t=0}) reduces to Eq.~(\ref{g-factor-eff-1}). At high 
fields, where $|g_{\rm eff}| \mu_B B \gtrsim |E_{lh1} - E_{hh2}|$ but 
$|\Delta E_Z|$ is still smaller than the distance to other hole
subbands, the Zeeman splitting is proportional to
$\sqrt{B}$. Correspondingly, the dash-dot line in 
Fig.~\ref{fig:zeeman-num} demonstrates a clear square-root asymptotics.

The allowance for the heavy-light mixing at interfaces stabilizes the
linear variation of $\Delta E_Z$ 
with the magnetic field. In Fig.~\ref{fig:zeeman-num} we compare the
magnetic-field dependence of the spin splitting  
depicted in the linear approximation (dashed) with the result of numerical
calculation performed beyond the this approximation.
For illustration we chose two widths of the GaAs/AlAs QW,
$a=40$~\AA~and $a=50$~\AA.~At $a=40$~\AA~ the light hole $lh1$ lies lower
in energy than the heavy hole $hh2$ and the Zeeman splitting is positive. For
$a=50$~\AA~QW, the relative energy positions of the $hh2$
  and $lh1$ states reverse, and the 
sign of Zeeman splitting changes respectively. One can see from
Fig.~\ref{fig:zeeman-num} that the  
linear interpolation somewhat overestimates absolute values of the
Zeeman splitting. 

\section{Conclusion}\label{sec:conclusions}

We have demonstrated that the proximity of the lowest light-hole
($lh1$) and first excited heavy-hole ($hh2$) subbands is responsible for a
giant contribution to the Zeeman splitting of hole states. It is shown
that both the magnitude and the sign of hole effective $g$-factor are very
sensitive to the structure parameters, in particular, to the quantum
well width, barrier height and heavy-light hole interface mixing
parameter. We have analyzed the Zeeman splitting of light holes in
a wide range of magnetic fields and derived equations for the differential
magnetoabsorption and magnetotransmission spectra with allowance for the
$lh1$-$hh2$ mixing.

Authors are grateful to E.Ya. Sherman for useful discussions. This work was financially supported by RBFR, ``Dynasty'' Foundation--ICFPM and EU projects Spinoptronics and POLAPHEN.


\begin{thebibliography}{10}
%\providecommand{\selectlanguage}[1]{\relax}
%%\input{babelbst.tex}
%\newcommand{\Capitalize}[1]{\uppercase{#1}}
%\newcommand{\capitalize}[1]{\expandafter\Capitalize#1}

\bibitem{PhysRev.114.90}
L.~M. Roth, B.~Lax, S.~Zwerdling,
%\newblock \emph{Theory of optical magneto-Absorption Effects in
%  Semiconductors}.
Phys. Rev. 114 (1959) 90.

\bibitem{ivchenko_kiselev92}
E.~L. Ivchenko, A.~A. Kiselev,
%\newblock \emph{Электронный g-фактор в квантовых ямах и сверхрешетках}.
Fiz. Tekh. Poluprovodn. 26 (1992) 1471 (1992) (trasl: Sov.
Phys. Semicond. 26 (1992) 827).

\bibitem{PhysRevB.75.245302}
I.~A. Yugova, A.~Greilich, D.~R. Yakovlev, A.~A. Kiselev, M.~Bayer, V.~V.
  Petrov, Y.~K. Dolgikh, D.~Reuter, A.~D. Wieck,
%\newblock \emph{Universal behavior of the electron $g$ factor in
%  $\mbox{GaAs/Al}_x\mbox{Ga}_{1-x} \mbox{As}$ QWs}.
Phys. Rev. B 75 (2007) 245302.

\bibitem{ivchenko05a}
E.~L. Ivchenko,
Optical Spectroscopy of Semiconductor Nanostructures, Alpha
Science Internat., Harrow, UK, 2005.
\bibitem{efros1} Th. Wimbauer, K. Oettinger, Al.L. Efros, B.K. Meyer, H. Brugger, Phys. Rev. B 50 (1994) 8889.
\bibitem{kiselevmoiseev96_rus}
A.A. Kiselev, L.V. Moiseev,
%\newblock \emph{Zeeman splitting of heavy-hole states in III-V and II-VI heterostructures}.
Fiz. Tverd. Tela 38 (1996) 1574 (transl: Phys. Solid State 38 (1996) 866).
\bibitem{kiselev_inplane} A.A. Kiselev, K.W. Kim, E. Yablonovich, Phys. Rev. B 64 (2001) 125303.
\bibitem{carmel} O. Carmel, H. Shtrikman, I. Bar-Joseph, Phys. Rev. B 48 (1993) 1955.
\bibitem{efros2} D.M. Hofmann, K. Oettinger, Al.L. Efros, B.K. Meyer, Phys. Rev. B 55 (1997) 9924.
\bibitem{10.1063/1.2245213}
Y.H. Chen, X.L. Ye, B.~Xu, Z.G. Wang, Z.~Yang,
%\newblock \emph{Large g factors of higher-lying excitons detected with
%  reflectance difference spectroscopy in GaAs-based QWs}. 
Appl. Phys. Lett. 89 (2006) 051903.
\bibitem{PhysRevB.82.195317}
P.~Renucci, V.G. Truong, H.~Jaffr\`es, L.~Lombez, P.H. Binh, T.~Amand, J.M.
 George, X.~Marie.
%\newblock \emph{Spin-polarized electroluminescence and spin-dependent
%  photocurrent in hybrid semiconductor/ferromagnetic heterostructures: An
%  asymmetric problem}.
Phys. Rev. B 82 (2010) 195317.
\bibitem{birpikusbook}
G.L. Bir, G.E. Pikus,
Symmetry and strain-induced effects in semiconductors, Nauka, Moscow, 1972; Wiley, New York,
1974.
\bibitem{Mar99}
X.~Marie, T.~Amand, P.~Le~Jeune, M.~Paillard, P.~Renucci, L.E. Golub, V.D.
Dymnikov, E.L. Ivchenko,
%\newblock \emph{Hole spin quantum beats in quantum-well structures}.
Phys. Rev. B 60 (1999) 5811.
\bibitem{PhysRev.102.1030}
J.~M. Luttinger,
%\newblock \emph{Quantum Theory of Cyclotron Resonance in Semiconductors:
%  General Theory}.
Phys. Rev. 102 (1956) 1030.
\bibitem{PhysRev.133.A542}
L.M. Roth,
%\newblock \emph{Theory of the Faraday Effect in Solids}.
Phys. Rev. 133 (1964) A542.
\bibitem{PhysRevB.83.165450}
P.~Moon, W.J. Choi, J.D. Lee.
%\newblock \emph{Electrically driven singularity and control of carrier spin of
 % a hybrid QW}.
Phys. Rev. B 83 (2011) 165450.
\bibitem{Rashba1988175}
E.~I. Rashba, E.~Y. Sherman,
%\newblock \emph{Spin-orbital band splitting in symmetric QWs}.
Physics Letters A 129 (1988) 175.
\bibitem{aleiner92}
I.~L. Aleiner, E.~L. Ivchenko,
%\newblock {\selectlanguage{russian}\emph{Природа анизотропного обменного
%  расщепления в сверхрешетках $\mbox{GaAs/AlAs}$ типа $\mbox{II}$}}.
Pis'ma Zh. Eksp. Teor. Fiz. 55 (1992) 662 (transl: JETP Letters 55 (1992) 692).
\bibitem{ivchenko96}
E.~Ivchenko, A.~Kaminski, U.~Roessler,
%\newblock \emph{Heavy-light hole mixing at zinc-blende (001) interfaces under
%  normal incidence}.
Phys. Rev. B 54 (1996) 5852.
\bibitem{toropov:035302}
A.~A. Toropov, E.~L. Ivchenko, O.~Krebs, S.~Cortez, P.~Voisin, J.~L. Gentner,
%\newblock \emph{Excitonic contributions to the quantum-confined Pockels
%  effect}.
Phys. Rev. B 63 (2001) 035302.
\bibitem{vurgaftman02}
I.~Vurgaftman, J.R. Meyer, L.R. Ram-Mohan,
%\newblock \emph{Band parameters for III--V compound semiconductors and their
%  alloys}.
J. Appl. Phys. 89 (2001) 5815.
\bibitem{FTP_1988} E.L. Ivchenko, P.S. Kop'ev, V.P. Kochereshko, I.N. Uraltsev, D.R. Yakovlev,
S.V. Ivanov, B.Ya. Meltzer, M.A. Kaliteevskii, Fiz. Tverd. Tela 22 (1988) 784 (transl: Sov. Phys. Semicond. 22 (1988)
497).
\bibitem{bychkov84}
Y.~Bychkov, E.~Rashba,
%\newblock \emph{Oscillatory effects and the magnetic susceptibility of carriers
%  in inversion layers}.
J. Phys. C: Solid State 17 (1984) 6039.

\bibitem{RevModPhys.82.3045}
M.Z. Hasan, C.L. Kane,
%\newblock \emph{\textit{Colloquium}: Topological insulators}.
Rev. Mod. Phys. 82 (2010) 3045; B. Buttner, C. X. Liu, G. Tkachov, E. G. Novik, C.~Brune, H.~Buhmann, E. M. Hankiewicz, P.~Recher, B.~Trauzettel, S. C. Zhang and L.W. Molenkamp, Nature Physics (2011) 1914.

\bibitem{rashba64}
E.I. Rashba, 
%\newblock \emph{Properties of semiconductors with an extremum loop. I.
%  Cyclotron and comninational resonance in a magnetic field perpendicular to
%  the plane of the loop}.
Sov. Phys. Solid State 2 (1964) 1874.

\bibitem{PhysRevLett.96.086805}
E.~McCann, V.~I. Fal'ko,
%\newblock \emph{Landau-Level Degeneracy and Quantum Hall Effect in a Graphite
%  Bilayer}.
Phys. Rev. Lett. 96 (2006) 086805.

\bibitem{Averkiev2005543}
N.~Averkiev, M.~Glazov, S.~Tarasenko,
%\newblock \emph{Suppression of spin beats in magneto-oscillation phenomena in
%  two-dimensional electron gas}.
Solid State Commun. 133 (2005) 543.

\bibitem{PhysRevB.37.4171}
P.~Lefebvre, B.~Gil, J.P. Lascaray, H.~Mathieu, D.~Bimberg, T.~Fukunaga,
  H.~Nakashima,
%\newblock \emph{Magnetoexcitons in a narrow single
%  GaAs-${\mathrm{Ga}}_{0.5}$${\mathrm{Al}}_{0.5}$As QW grown by
%  molecular-beam epitaxy}.
Phys. Rev. B 37 (1988) 4171.

\bibitem{PhysRevB.48.1955}
O.~Carmel, H.~Shtrikman, I.~Bar-Joseph,
%\newblock \emph{Quantum-beat spectroscopy of the Zeeman splitting of heavy- and
%  light-hole excitons in
%  GaAs/${\mathrm{Al}}_{\mathit{x}}$${\mathrm{Ga}}_{1\mathrm{-}\mathit{x}}$As
%  QWs}.
Phys. Rev. B 48 (1993) 1955.

\end{thebibliography}
\end{document}